\documentclass[aps,pre,twocolumn,showpacs,amssymb,groupedaddress]{revtex4-1}
\usepackage{graphicx}
\usepackage{amsmath}

\begin{document}

% Use the \preprint command to place your local institutional report
% number in the upper righthand corner of the title page in preprint mode.
% Multiple \preprint commands are allowed.
% Use the 'preprintnumbers' class option to override journal defaults
% to display numbers if necessary
%\preprint{}

%Title of paper
\title{Spatial distribution functions of random packed granular spheres obtained by direct particle imaging}
\author{Andreea Panaitescu and Arshad Kudrolli}

\affiliation{Department of Physics, Clark University, Worcester, MA 01610}

\date{(\today)}

\begin{abstract}
We measure the two-point density correlations and Voronoi cell distributions of cyclically sheared granular spheres obtained with a fluorescence technique and compare them with random packing of frictionless spheres. We find that the radial distribution function $g(r)$ is captured by the Percus-Yevick equation for initial volume fraction $\phi=0.59$. However, small but systematic deviations are observed because of the splitting of the second peak as $\phi$ is increased towards random close packing. The distribution of the Voronoi free volumes deviates from postulated $\Gamma$ distributions, and the orientational order metric $Q_6$ shows disorder compared to numerical results reported for  frictionless spheres. Overall, these measures show significant similarity of random packing of granular and frictionless spheres, but some systematic differences as well.
\end{abstract}

% insert suggested PACS numbers in braces on next line

\pacs{45.70.Qj, 05.65.+b}

% insert suggested keywords - APS authors don't need to do this
%\keywords{}

%\maketitle must follow title, authors, abstract, \pacs, and \keywords

% References should be done using the \cite, \ref, and \label commands

% insert suggested keywords - APS authors don't need to do this
%\keywords{}

%\maketitle must follow title, authors, abstract, \pacs, and \keywords
%\pacs{45.70.Mg}
\maketitle

% Refer
%\section{Introduction}

The packing of spheres is one of the enduring problems in physics, and a basis to understand the structure and strength of granular matter. Assuming dominance of steric interactions, dense packing of steel spheres was first used to understand structure of simple liquids with the radial distribution function $g(r)$ and the orientation order metric $Q_6$~\cite{scott62}. However, experimental measurements at boundaries~\cite{mueth98} and computer simulations in the bulk~\cite{silbert02} have since shown that inter-particle friction can affect granular packing. %The nature of granular packing remains an open question for several reasons. 
Friction between particles changes the fundamental stability condition at contact from the frictionless case, causing a packing to be protocol dependent and the system to be out-of-equilibrium. 

The difficulty of accurately measuring significant number of particle positions in the bulk away from the influence of boundaries has also stymied progress. Recent experimental studies~\cite{aste08,losert} have examined packing of granular spheres and find that the associated free volume distributions are described by a $\Gamma$ distribution with two fitting parameters which were then given a thermodynamic interpretation~\cite{aste08}. These results are puzzling in light of earlier analytical work in one-dimension and simulations in two and three dimensions that show a $\Gamma$ distribution with 3 fitting parameters is needed to describe a broad range of volume fraction for elastic particles~\cite{kumaran05}. 

Here, we discuss new experiments with spherical granular particles which enable us to directly determine statistical measures to understand the effect of friction, test the effect of shear, and perform a rigorous comparison with frictionless hard sphere packing. Using a fluorescence technique~\cite{tsai03,siavoshi06,losert}, we obtain the packing of glass spheres before and after application of cyclic shear, and compare with random packing of frictionless spheres. We find that the overall shape of $g(r)$ for volume fraction $\phi \sim 0.6$ is captured by the Percus-Yevick equation~\cite{percus57} which assumes random packing of spheres without angular correlations. But, systematic deviations are observed because of the splitting of the second peak as $\phi$ is increased toward random close packing, $Q_6$ shows partial hexagonal order, and the distribution of the Voronoi free volumes shows enhanced probabilities at higher values compared with $\Gamma$ distributions postulated~\cite{kumaran05} for random packing of spheres.

%\section{Experimental Procedures and Apparatus}
\begin{figure}
\includegraphics[width=0.85\linewidth]{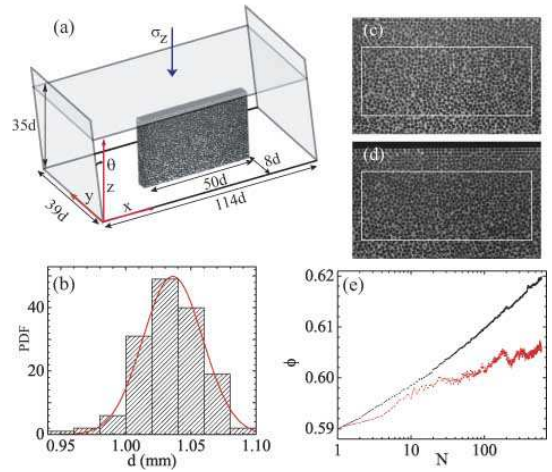}
\caption{(a) Schematic diagram of the cyclic shear cell used in the  experiments.  (b) An image of the initial packing observed in the central vertical slice of the shear cell after particles are filled inside the cell. (c) Ordering grows near the top boundary which is free to move after 600 shear cycles but the packing in the bulk appears random. (d) The probability distribution function of the diameter of the glass beads. (e) The volume fraction $\phi$ as a function of shear cycle number measured in the bulk (red/grey) evolves more slowly than in the entire cell (black) because of the ordering near the boundaries. }
\label{schem}
\end{figure}

The experiments to measure structure of granular packing are performed using a shear cell shown schematically in Fig.~\ref{schem}(a). The parallelepiped shaped cell consists of a rigid front, back, and bottom transparent glass boundary, and side boundaries that can be tilted through a prescribed angle $\theta$ to cyclically shear and perturb the packing.  Glass spheres with density  $\rho_g = 2.5 \times 10^{3}$ kg m$^{-3}$ and average diameter $d = 1.034$\,mm are gently added inside a shear cell filled with an interstitial liquid with the same refractive index as the glass spheres, density $\rho_l =  1.0 \times 10^{3}$ kg m$^{-3}$, and viscosity $\nu = 2.2 \times 10^{-2}$ Pa\,s.   Then a flat plate is placed on the top, which is constrained to move only in the vertical and horizontal direction and not allowed to rotate using a rigid set of linear guides. The initial volume fraction of the glass beads is measured using the mass of the particles added and the volume of the cell occupied and found to be $\phi = 0.59$. A normal stress $\sigma_z = 0.4$\,Pa is applied on the top boundary which is about five times the net gravitational stress due to the weight of the grains alone inside the cell, and is found to eliminate gradients due to gravity in the system. 

A dye is added to the liquid and a thin slice of the cell is illuminated with a laser and a cylindrical lens~\cite{orpe07}. The resulting fluorescent light causes the particles to appear dark against a bright background, and is imaged from an orthogonal direction with a resolution of 20 pixels to a particle diameter using a $1000 \times 1000$ pixel 10-bit camera. A stack of images is recorded by linearly translating the plane of illumination. We then examine a $40d \times 5d \times 17.5d$ central region as in indicated by box in Fig.~\ref{schem}(b,c) to avoid any effect of the boundaries, and locate the absolute position of the spheres to within the slight polydispersity of the particles (see Fig.~\ref{schem}(d)).  

We impose quasi-static shear strain by varying $\theta$ between $\pm \pi/36$ radians with a mean angular speed $\omega = 8.0 \times 10^{-3}$rad s$^{-1}$, with a wait time of 50\,s while the stack of images is acquired every time the system returns to its original position,  $\theta_0 = 0$\,rad. The lubrication forces~\cite{pitois} due to liquid draining at contacts can be estimated to be $10^{-5}$ lower than the confining forces, and the particle Reynolds number is $\sim 10^{-1}$. Therefore the particles can be assumed to be in contact during the entire experiment and the interstitial liquid does not have any impact on the observed structure. 

The mean packing volume fraction of the spheres inside the entire cell is first obtained by measuring the position of the top plate as a function of the shear cycles. The mean $\phi$ is observed to increase logarithmically from the initial value by 5\% (Fig.~\ref{schem}e) consistent with previous reports with a similar setup~\cite{pouliquen}. However, it is noteworthy that this is the total $\phi$ inside the cell and can be different than $\phi$ in the bulk because of influence of boundaries~\cite{orpe07}. Examining the images corresponding to the initial state of the packing, before applying the shear deformation, $N = 1$ and after shear cycle $N = 600$ shown in Fig.~\ref{schem}(b,c), we indeed note greater ordering near the top where the boundary shears the spheres and moves to accommodate changes in the total $\phi$. The boundary between ordered and disordered region appears sharp and moves downward as $N$ is increased further, similar to development of crystalline order inside a Couette shear cell upon extended shearing~\cite{tsai03}. Therefore, we focus on the first $600$ shear cycles where the particles inside the bulk in the observation window appear uniformly random and obtain $\phi$ from the ratio of the particle volume and the average Voronoi volume in the bulk. The Voronoi volume is defined by points in the volume closest to that particle, and is calculated using algorithms written by Rycroft~\cite{rycroft} and measured particle positions. As shown in Fig.~\ref{schem}(e), the evolution of $\phi$ in the bulk is observed to be slower compared with $\phi$ measured for the entire cell and is used in all subsequent discussion.

%\section{Radial distribution function}

\begin{figure}
\includegraphics[width=0.75\linewidth]{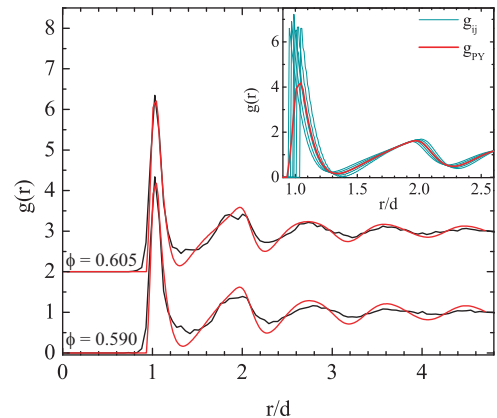}
\caption{The radial distribution function $g(r)$ plotted as function of distance  $r$  normalized by the mean diameter $d$ for initial packing $\phi = 0.59$ and final packing $\phi = 0.605$ obtained after shearing, (black), is compared with the theoretical calculation (red/grey) obtained by using the Percus-Yevick equation, and the measured polydispersity of the beads. The $\phi = 0.605$ case is offset for clarity. Inset: The calculated pair distribution functions $g_{ij}$ for particles coarsened to three sizes ($\phi = 0.59$). The thick red/grey curve represents the average of the six distinct contributions $g_{ij}(r)$ and is used for comparison with the experimental g(r).}
\label{gofr}
\end{figure}

To analyze the structure from the measured particle positions, we first discuss the radial distribution function $g(r)$ which represents the probability that the center of a particle is found at a distance $r$ from another particle.  
$g(r)$ obtained from the experimental data is shown in Fig.~\ref{gofr}. For the initial volume fraction $\phi = 0.59$, $g(r)$ shows a tall peak at $r \sim d$, and a broad peak at $r \sim 2d$, but for the higher $\phi$ obtained after cyclic shear, the second peak splits and a weak secondary peak occurs at $r = \sqrt{3}d$, corresponding to the next nearest neighbor in a face-centered-cubic (FCC) lattice. 

Now, the Percus-Yevick equation~\cite{percus57,pv} can be used to analytically find $g(r)$ for randomly distributed particles for a given particle size, and volume fraction. Because of the slight polydispersity of the particles used in our experiments, we in fact have to compute the Percus-Yevick pair distribution function for polydisperse spherical packings $g_{ij}(r)$, where the indices $i,j$ represent the probability of finding a particle with diameter $d_j$ at a distance $r$ from a particle with diameter $d_i$. 
\begin{equation} \label{gr_PY}
g_{PY}(r) = \frac{1}{n(n-1)}\sum_{\substack{i=1 \\ {j\ge i}}}^{n}g_{ij}(r)
\end{equation}
here $n$ is the number of particle sizes in the system. Coarse graining the observed size distribution - shown in Fig.~\ref{schem}(d) - into three sizes $d_1 = 0.98$\,mm, $d_2 = 1.02$\,mm, $d_3 = 1.07$\,mm (using a greater $n$ does not change the results significantly), we calculate the corresponding six distinct $g_{ij}(r)$ terms, which are plotted in the Inset to Fig.~\ref{gofr} for $\phi = 0.59$, and thus the computed $g_{PY}(r)$ for the polydisperse packing according to the Percus-Yevick approximation, which is plotted in Fig.~\ref{gofr}. We observe that the amplitude and width of the primary peak and the broad features of the secondary peaks are in good agreement with the Percus-Yevick approximation. This comparison is especially noteworthy because there are no fitting parameters. At higher $\phi$, the primary peak and the overall form is still captured by the $g_{PY}(r)$, but details such as the splitting of the second peak which can indicate hexagonal ordering is not captured because Percus-Yevick approximation assumes random angular orientation. 

%\section{Bound orientational order parameter}

\begin{figure}
\includegraphics[width=0.75\linewidth]{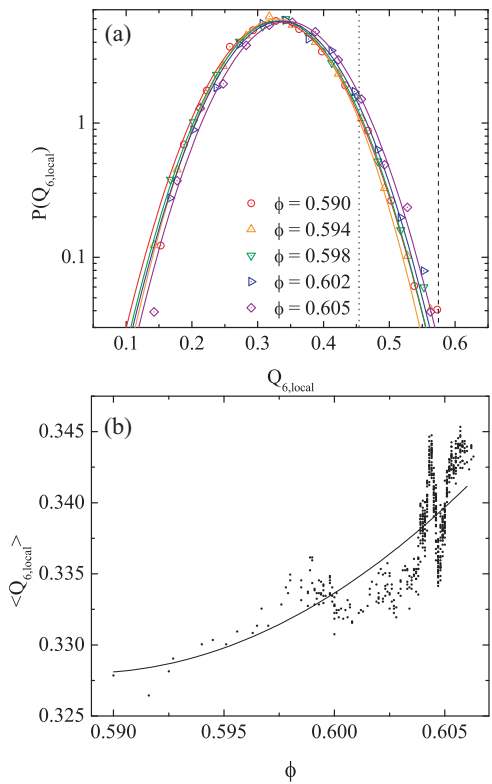}
\caption{(a) The probability distribution of $Q_6$ measured for each particle using Voronoi neighbors are observed to described by Gaussian fits. (b) The mean $Q_{6,local}$ measured as a function of $\phi$. The curve is a guide to the eye and shows that $Q_6$ increase somewhat over the narrow $\phi$ investigated. % I am unsure if (b) is necessary because it really makes clear we didn't do a large range of phi. 
}
\label{Q6}
\end{figure}

To investigate orientational order in the packing, we use the orientation order metric
\begin{equation} \label{q6}
Q_{l} \equiv \sum_{i=1}^{N} \left( \frac{4\pi}{l(l+1)} \sum_{m=-l}^{m=l} {\left|\langle Y_{lm}(\Theta({\bf r}),\Phi({\bf r}) )\rangle \right|}^2 \right)^{1/2},
\end{equation} 
where, $l = 6$ to examine hexagonal order, $Y_{lm}$ are the spherical harmonics, $\Theta({\bf r})$ and $\Phi({\bf r})$ are the polar and azimuthal angle, respectively, and {\bf r} is the position vector from a particle to its neighbor~\cite{Steinhardt}. We define particle neighbors as those which share a Voronoi cell surface~\cite{rycroft}. This removes any ambigity as is introduced when considering only neighbors at contact due to roundoff errors in finding a particle center. In order to compare with packing of elastic particles, we first compute $Q_{6,global}$ by averaging $Y_{lm}(\Theta({\bf r}),\Phi({\bf r}))$ over all the bonds of the packing, and find $Q_{6,global} = 0.27 \pm 0.02$. If particles neighbors are uncorrelated, then $Q_6$ is small because it goes as square root of the total number of bonds~\cite{rintoul}. On the other hand, $Q_6$ for a FCC crystal with 12 neighbors is 0.5745. But even a slight perturbation due to roundoff errors introduces 2 extra neighbors and the corresponding $Q_6$ is on average 0.454~\cite{gervois}. Therefore, the observed distribution shows ordering but the degree of order appears lower compared with simulations of frictionless hard spheres~\cite{kumaran06}, where $Q_6$ as high as 0.4 was reported for a frictionless hard spheres at comparable $\phi$. In those studies which were performed with considerably smaller system size, particle inelasticity was observed to lower $Q_6$ but not as significantly as in our experiments. 

To examine the local orientational order more closely, we plot the observed probability distribution of $Q_6$ for each particle in Fig.~\ref{Q6}(a), and the mean of the distribution $\langle Q_{6,local} \rangle$ in Fig.~\ref{Q6}(b). $\langle Q_{6,local} \rangle$ is more sensitive than $Q_{6,global}$ to small crystalline regions within a packing and allows us to avoid the possibility of destructive interference between different crystalline regions~\cite{kansal}. No significant enhancement of distribution is found at the values corresponding to FCC crystal, and the observed $Q_6$ distribution can be described rather well in fact by Gaussian fits. From these observations we conclude that while there is some local hexagonal order which increases slightly over the $\phi$ investigated, no significant crystallites occur in this dense regime approaching random close packing.

%\section{Voronoi tessellation}

\begin{figure}
\includegraphics[width=0.75\linewidth]{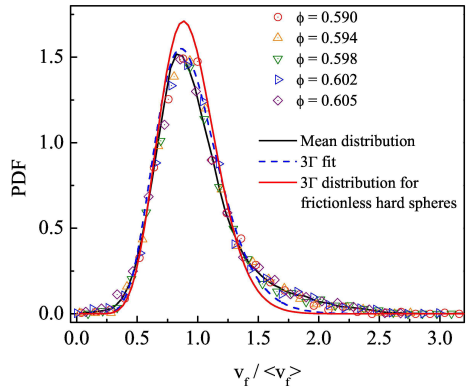}
\caption{The probability distribution function of the free volume associated with a sphere $v_f$ normalized by the mean free volume $\langle v_f \rangle$ plotted for various $\phi$. The smooth black curve is obtained after averaging over all the experimental data. $\Gamma$ function corresponding to elastic frictionless spheres, the smooth red/grey curve, is shown for comparison, and is observed to deviate systematically at higher $v_f$. Allowing the fitting parameters to float  improves the fit, but systematic differences persist for $v_f > \langle v_f \rangle$ (blue dashed curve).} 
\label{voronoi}
\end{figure}

A complementary method to examine the packing at the particle scale is using the free volume $v_f$ associated with each particle given by subtracting the minimum Voronoi volume corresponding to close packing, $v_c = d^3/\sqrt{2}$ from the Voronoi volume. This statistical quantity has gained prominence because it may  be used to define a new measure of entropy based on disorder in packing~\cite{kumaran05,briscoe08}, and may be amenable to thermodynamic interpretation~\cite{aste08}.   It has been postulated based on analytical work in 1-dimensional systems, that $v_f$ distribution of random packing of spheres can be described by a $\Gamma$ distribution~\cite{kumaran05}:
\begin{equation}
f(v_f) = \frac{\delta\alpha^{(m/\delta^2)}}{\Gamma(m/\delta^2)}{v_f}^{(m/\delta - 1)}e^{-\alpha{v_f}^\delta}
\label{3gamma}
\end{equation}
with three fitting parameters $m$, $\delta$ and $\alpha$ that control different parts of the distribution and were determined by numerical simulations with frictionless hard spheres~\cite{kumaran05}. 
In Fig.~\ref{voronoi}, we plot $v_f$ normalized by the mean free volume $\langle v_f \rangle$ at that $\phi$ along with the mean distribution obtained after averaging over all the measured $\phi$. The errors associated with the slightly polydispersity and errors in finding particle centers is of order of symbol size. Further, we plot Eq.~\ref{3gamma} using $m = 15.5$, $\delta = 1.3$ reported in Ref.~\cite{kumaran05}, and allowing $\alpha$ to float to obtain best fit.  Clearly, systematic deviations are observed from the frictionless case, complementing the results for $Q_6$.

In order to check if a $\Gamma$-distribution can capture the experimentally observed free volume distributions, we tested both the three fitting parameter distribution, and the two fitting parameter distribution, which corresponds to setting $\delta = 1$ in Eq.~\ref{3gamma}. The best fit obtained with $m = 12.3$, $\alpha = 24.5$, and $\delta = 0.73$ is also shown in Fig.~\ref{voronoi}. Even in this case we obtain enhanced probabilities for $v_f$ greater than the mean. Therefore, our distribution differ from the experimental distributions used to give a simple thermodynamic interpretation of granular packing~\cite{aste08}. While it is possible that such deviations arise because of the differences in preparation protocol, we note no significant differences in the distributions obtained before and after application of cyclic shear in our experiments.

In conclusion, we measured packing of granular spheres and compared experimentally obtained two-point density correlations and free volume distributions before and after application of shear.  We find that the radial distribution function is captured overall by the Percus-Yevick equation, which is important because it is fundamental to calculating the strength, heat conduction, and electro-magnetic wave scattering properties of a material. However, angular correlation can be observed using the orientational order metric. In comparing with numerical simulations reported for frictionless sphere at comparable volume fractions, we find systematic differences in packing as measured by lower angular correlations, and deviation of free volume distributions from $\Gamma$ distributions postulated for frictionless hard spheres.  

\begin{acknowledgments}
We thank V. Kumaran, Ashish Orpe, and Michael Berhanu for many stimulating  discussions, and Chris Rycroft for providing us Voro++ software library. This work was supported by the National Science Foundation under NSF Grant No CBET 0853943. 
\end{acknowledgments}

% Create the reference section using BibTeX:
\bibliographystyle{apsrev}
%\bibliography{ak-bib}

\end{document}